\documentclass[arxiv, twocolumn, superscriptaddress]{revtex4}

\usepackage{amsfonts}           % AMS packages
\usepackage{amssymb}            % ^
\usepackage{amsmath}            % ^
\usepackage{graphicx}   % For production with eps / jpg figures
\usepackage{enumerate}
\usepackage{bm}                 % A better "bold math" package -- any math symbol can be emboldened with the command \bm{*}
\usepackage[ulem=normalem]{changes}
\newcommand{\bea}{\begin{eqnarray}}
\newcommand{\eea}{\end{eqnarray}}

\begin{document}

%Title.  Two "affiliation" class options are groupedaddress (default) and superscriptaddress
\title{Direct detection of plasticity onset through total-strain profile evolution}
\date{\today}
%key words: power law statistics, exponent, event spacial distribution, dislocation drag. 
\author{Stefanos Papanikolaou}
\affiliation{NOMATEN Centre of Excellence, National Centre of Nuclear Research, A. Soltana 7, 05-400 Otwock–\'Swierk, Poland.}
\author{Mikko J. Alava}
%\affiliation{Department of Mechanical Engineering, Johns Hopkins University, Baltimore, MD21218, United States.}
\affiliation{NOMATEN Centre of Excellence, National Centre of Nuclear Research, A. Soltana 7, 05-400 Otwock–\'Swierk, Poland.}
\affiliation{Department of Applied Physics, Aalto University,P.O. Box 11100, FI-00076 Aalto, Espoo, Finland.}
%key words: avalanche, strain rate effect, size effect, drag effect, power law distribution, events spacial distribution, events temporal distribution.  

\begin{abstract}
{  Plastic yielding in solids strongly depends on various conditions, such as temperature and loading rate and indeed, sample-dependent knowledge of yield points in structural materials promotes reliability in mechanical behavior. Commonly, yielding is measured through controlled mechanical testing at small or large scales, in ways that either distinguish elastic (stress) from total deformation measurements, or by identifying plastic slip contributions. In this paper we argue that instead of separate elastic/plastic measurements, yielding can be unraveled through statistical analysis of total strain fluctuations during the evolution sequence of profiles measured \emph{in-situ}, through digital image correlation. We demonstrate two distinct ways of precisely quantifying yield locations in widely applicable crystal plasticity models, that apply in polycrystalline solids, either by using principal component analysis or discrete wavelet transforms. We test and compare these approaches in synthetic data of polycrystal simulations  and a variety of yielding responses, through changes of the applied loading rates and the strain-rate sensitivity exponents.}
\end{abstract}

\maketitle 

%\section{Introduction}
%Technology --> Large Data --> New tools --> Multiple scales
The understanding and classification of complex patterns that emerge in science and engineering is a key component of technological developments across fields, introducing new dimensional reduction of tools and paradigms that perform at multiple scales. The basic constituent of such classification is a pattern characterization parameter that should be directly introduced from input data through simple computational steps. In the context of materials informatics {  for mechanical deformation applications, a major goal} is the information hidden in total strain maps, readily measured through digital image correlation~\cite{Schreier:2009xd}(DIC) {  consisting of surface texture changes. In crystal plasticity, information extracted from DIC has been focused on plastic slip accumulation in particular crystallographic directions, as well as crack initiation locations, always assisted by intense finite element modeling~\cite{roux, tarleton, daly}. However, this information is dependent on phenomenological constitutive laws that are changing with temperature and loading rate, especially under extreme conditions. In this paper, we introduce two functional tools, one based on principal component analysis and another on discrete wavelet transforms, that aim to capture the onset of crystal plasticity through the analysis of total strain fluctuations as they develop during mechanical loading. As a test case, we demonstrate how these tools can be implemented directly on DIC data in polycrystalline samples, synthetically produced using a widely applicable phenomenological crystal plasticity model for pure $Al$.

%Elastic-Plastic change elusive in that its onset is engineering-based... violated at small scales... But, truthfully, the onset is characterized by plastic strain, which becomes non-zero and becomes on average comparable at yield.
The application of modern statistical methods to identify transitions between phases with well-defined order parameters in materials has become quite common~\cite{and1}. The recent theme, in physics and engineering communities is the application of machine learning methods, so that the identification of a precise characterization parameter is avoided, and automatic detection is achieved~\cite{p1,p2,p3,p4,p5,p6,p7,p8}. However, the absence of detailed knowledge may lead to overfitting artifacts and unsuccessful machine learning training. In this context, the use of unsupervised machine learning through principal component analysis (PCA) has been insightful~\cite{pca-1,pca-2,pca-3,pca-4,pca-rottler}.}
\begin{figure}[t]
\centering
\includegraphics[width=0.5\textwidth]{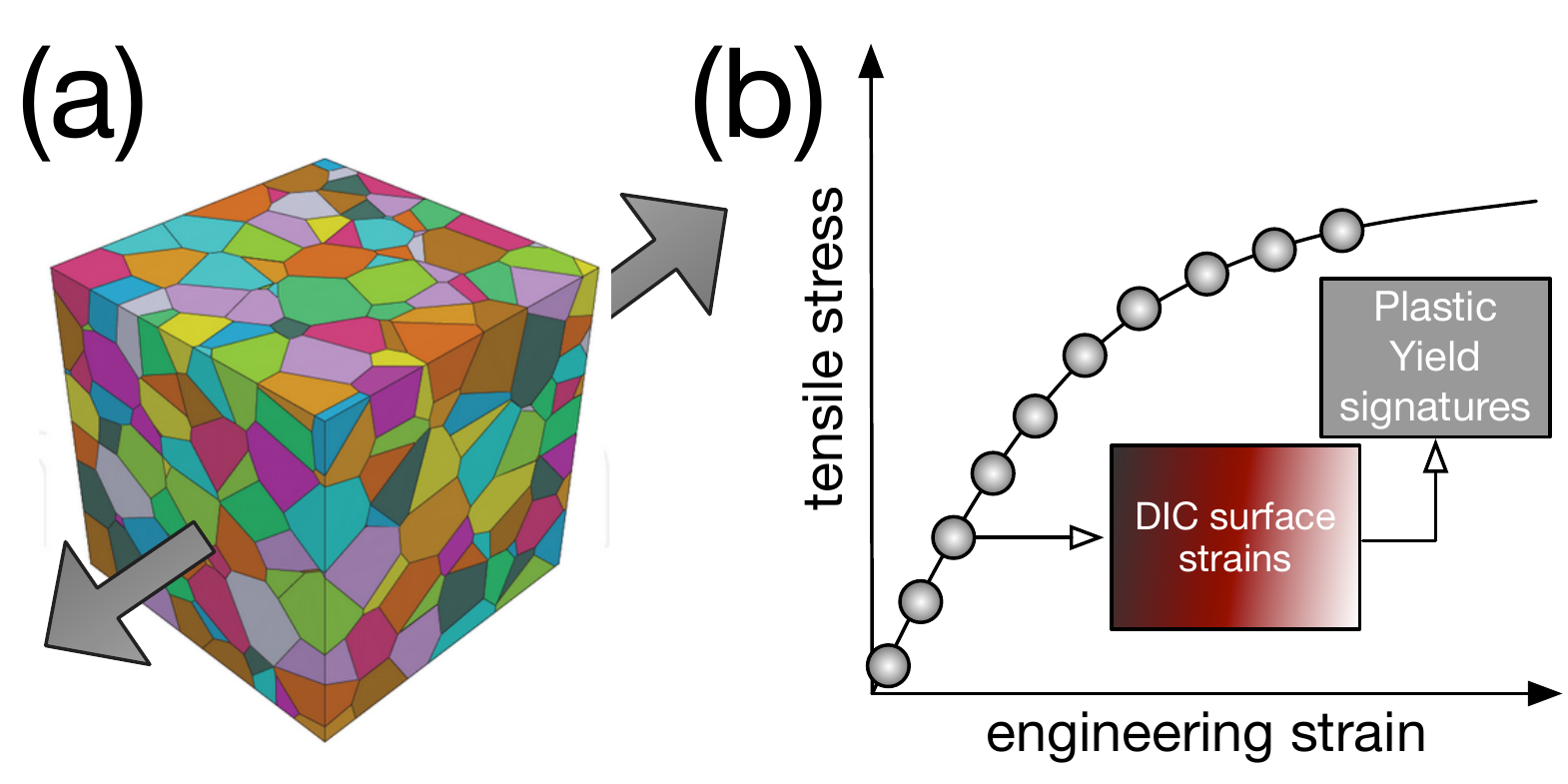}
\caption{{\bf Schematic of main scenario investigated in this work}. A polycrystalline sample with an unknown mechanical response is being uniaxially loaded under tension (a) and imaged on one of the surfaces through digital image correlation (DIC) (b). The total strain image output of DIC serves as the basis of an inquiry of yielding signatures in DIC correlations, as more images are collected. In this work, we utilize two novel tools based on principal component analysis (PCA) and discrete wavelet transforms (DWT). We consider various yield behaviors in this work by phenomenologically controlling the  loading rate and rate hardening exponent $(\dot\gamma,n)$.}
\label{fig:1}
\end{figure}

%The whole discussion of how one defines shear-induced yielding took another turn by the glass-community.
{  The sole investigation of total strains typically prohibits the identification of structural defects, such as dislocations in crystalline solids or structural hot spots in amorphous solids, since a separation of elastic and plastic contributions to the strain is required, if that defects carry plastic flow~\cite{richard2021}. In fact, typical stress-strain standard mechanical testing does precisely that kind of a separation, by plotting an elastic contribution (stress) as function of the total contribution (total strain) that includes the sum of elastic and plastic ones (for strains $<5\%$). However, the investigation of the onset of plasticity has always been connected to the daunting task of separation between elastic and plastic contributions that imposes the need to  perform either standard mechanical testing,
or non-destructive testing through nanoindentation and defect imaging~\cite{test-2}.  In any case, the determination of the yield stress of a material is non-trivial, depends on environmental conditions,  and there are still open questions in both  simulations and experiments, especially in the context of amorphous solids~\cite{glass1,glass2,glass3,glass4,glass5,glass6,glass7,glass8}.
}

\begin{figure*}[tbh]
\centering
\includegraphics[width=0.98\textwidth]{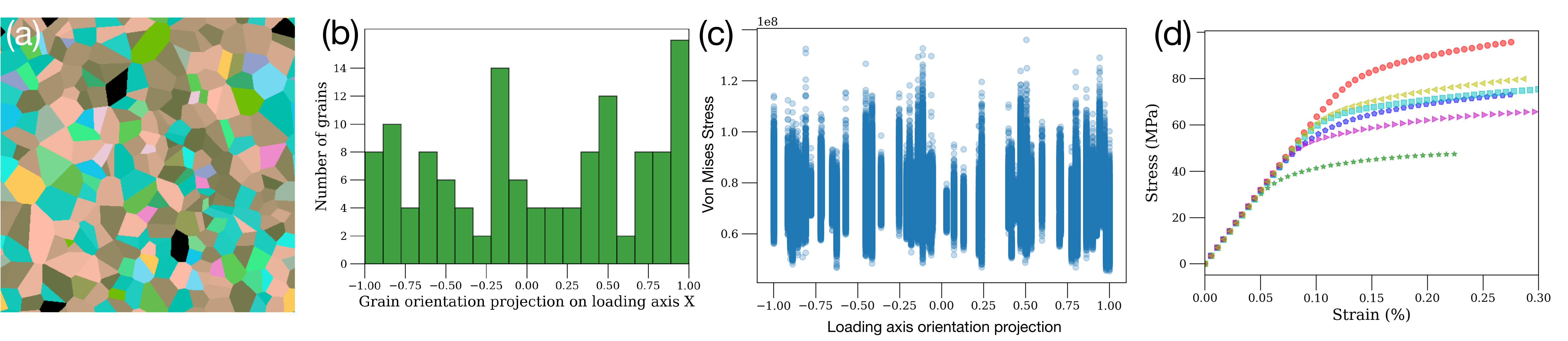}
\caption{{\bf Polycrystalline texture and mechanical response}. (a) Polycrystalline texture of samples studied in this work on 512x512x2 {  2D} and 64x64x64 {  (3D)} grids, with 256 grains in each microstructure, (b) Grain orientation distribution through loading-axis projection (given the representation of angles through Bunge Euler angles the projection on the x-axis is cos$\phi_1$cos$\phi_2$-sin$\phi_1$sin$\phi_2$cos$\Phi$) and (c) Von Mises stress values on the microstructure of the case $(\dot\gamma,n)=\{5\times10^{-2}/s,10\}$ {  for the 2D grid}, at total strain $0.25\%$. {  In (d) we show the mechanical response (Average stress along loading x-direction Vs. loading strain) for all studied samples, with loading along y-direction with various loading rates and rate sensitivity exponents $(\dot\gamma,n)=$ in the 2D:  $:\{5\times10^{-4}/s,5\}$, $\star:\{5\times10^{-4}/s,10\}$, $\bullet:\{5\times10^{-3}/s,5\}$, $:\{5\times10^{-3}/s,10\}$, $:\{5\times10^{-4}/s,10\}$  and also, 3D: $\square:\{5\times10^{-4}/s,10\}$. The 3D sample has identical loading and strain-rate sensitivity as the 2D case $:\{5\times10^{-4}/s,10\}$.}}
\label{fig:2}
\end{figure*}

%Recently, DIC methods have been quite simple to implement and proliferant.
{  Technological progress in image collection and analysis tools for materials science, have promoted DIC into an essential tool for acquiring full-field surface displacements towards strain maps at various resolutions, from mm to nm~\cite{dic1}. Through modern cameras and fast computational analysis, one may track displacements of distinct features in deformed samples. Nevertheless, DIC only tracks total displacements, and thus, its usefulness has mainly been the assessment of strain heterogeneity~\cite{dic2}, or to attempt an inverse solution by validating microstructurally accurate finite element modeling~\cite{dic3}. Indeed, DIC data has been useful for finding similarities to finite-element models of various types~\cite{dic3, Schreier:2009xd}, with advanced coarse-grained and multiscale models of microstructural deformation behavior~\cite{alava1,alava2,alava3}, and machine learning models~\cite{pap1,pap2,pap3}.  Nevertheless, the evolution of DIC total strain profiles contains enormous information that may be adequate for material properties related to yielding, hardening and eigenstrains. Here, we make a key initial step and demonstrate how to use strain maps to extract the yield point of a polycrystalline microstructure by using PCA or DWT.
}

%Question: How one can detect yielding directly from DIC output (total strain)?
In this paper, we focus on polycrystalline metals and small quasi-elastic strains ($<0.3\%$), and the incipient transition to plasticity. Our focus is on the development of tools that can reliably extract key material properties directly out of total strain maps, and fill two needs with one deed: perform efficient dimensional reduction of DIC image sequences, and also identify key material properties without the use of local or global stress information. For this purpose, we perform simulations in quasi-two-dimensional and three-dimensional samples, by using a phenomenological crystal plasticity model, that incorporates well established polycrystalline deformation mechanisms~\cite{roters2019}. In the context of this model, we demonstrate that PCA analysis of strain fluctuations can be used to accurately identify yield points. If we vary the mictrostructural hardening exponent and strain rate, the plasticity transition is qualitatively modified, but the developed measures still efficiently track the yield point. Furthermore, we  demonstrate that DWT coefficients, designed to identify localized features in images, can be used to identify the onset of plasticity at the material yield point. We conclude by comparing the two complementary approaches and discuss how to generalize them to learning additional properties, such as hardening and damage parameters, as well as mutually distinguish various hardening and damage mechanisms.

%The model
%\section{Results}
We consider synthetic data of uniaxial tensile mechanical deformation in a polycrystalline sample, where the physics of crystal plasticity is captured in the most common, recognizable way. 
\begin{figure*}[tbh]
\centering
\includegraphics[width=0.98\textwidth]{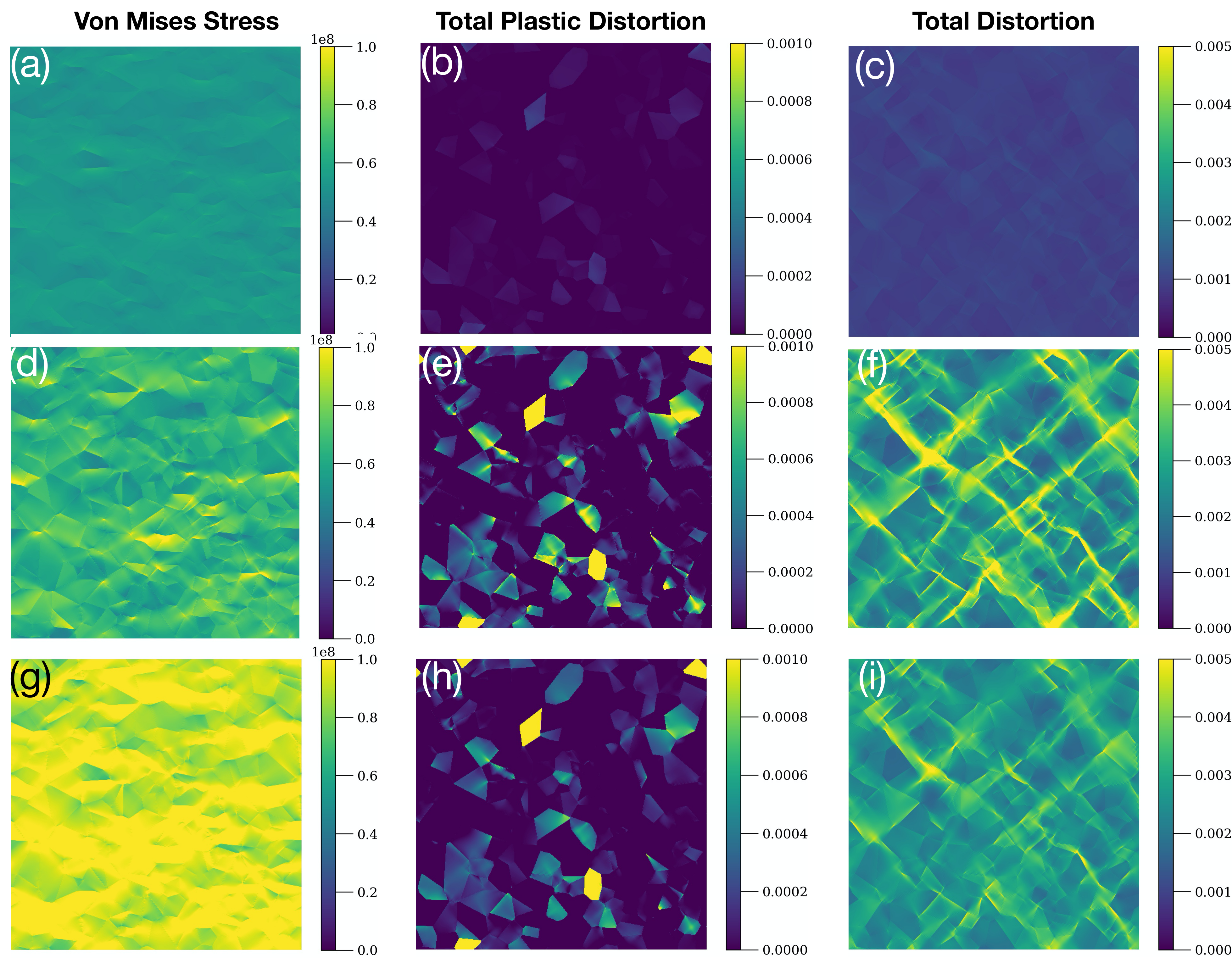}
\caption{{\bf Von Mises Stress, Plastic, and Total distortion  at the yield point and post-yield strain}. {  (a,d,g) Von Mises stress, (b,e,h) Plastic distortion defined as $|F_p-I|$, (c, f, i) }Total distortion, defined as $|F-I|$. (a,b,c) At yield point (0.12\% total strain), {  for 2D samples and} loading rate and rate sensitivity exponent $(\dot\gamma,n)=$ $\{5\times10^{-4}/s,10\}$, (d,e,f) at $0.25\%$ total strain, for loading rate and rate sensitivity exponent $(\dot\gamma,n)=$ $\{5\times10^{-4}/s,10\}$ (g,h,i) at $0.25\%$ total strain, for loading rate and rate sensitivity exponent $\{5\times10^{-2}/s,10\}$ }
\label{fig:3}
\end{figure*}
The model is uniaxially loaded along the x - direction, and x-y surface images were produced for close-by strain intervals up to, typically, $0.3\%$ total strain, where all samples have already yielded. The focus of this work is the analysis of these image sequences in a way that key information about the yielding transition is directly extracted from total-strain evolution (see also Fig.~\ref{fig:1}). We utilize phenomenological crystal plasticity in the continuum~\cite{asaro} to solve for material deformation due to elasticity, plasticity and damage evolution within the sample. 
\begin{figure*}[tbh]
\centering
\includegraphics[width=0.98\textwidth]{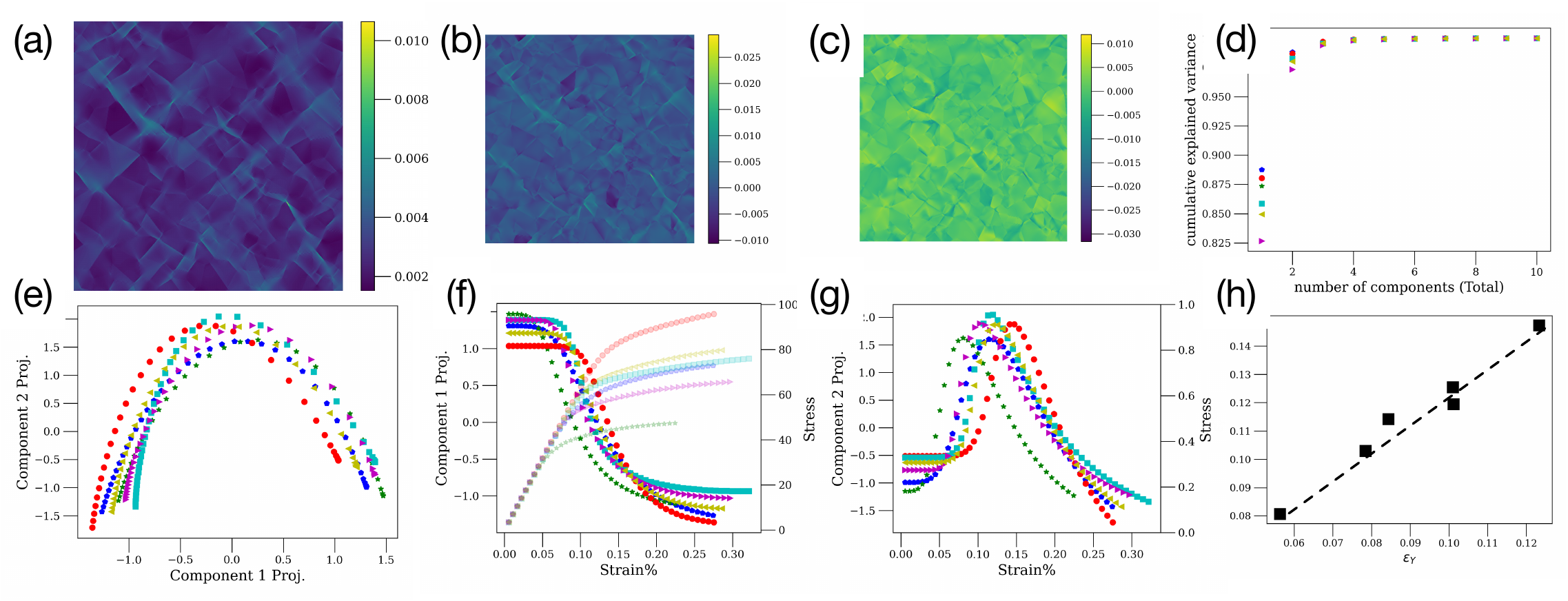}
\caption{
{\bf PCA of Total Strain Profiles and Order Parameters for Plasticity}: Principal component analysis is applied on properly normalized total strain ($|F-I|$) maps.    (a) Total strain map for strain $0.25\%$ and $(\dot\gamma,n)=$ $:\{5\times10^{-3}/s,10\}$, similar to the cases shown in Fig.~\ref{fig:3}(c,f,i) for examples), after subtracting the mean, taking the absolute value and then, dividing with the sample strain variance.
(b) First and (c) Second PCA component for $(\dot\gamma,n)=\{5\times10^{-4}/s,10\}$
 (d) Principal component cumulative variance of the components, showing a saturation to more than 99\% of the observed variability by just utilizing 2 components, 
(e) The projection of the first ($P^{(k)}_1$) and second ($P^{(k)}_2$) components on the strain map samples, after normalizing with $\sqrt{\sigma_i}$ where $\sigma_i$ is the corresponding singular case, for the various material cases discussed alongside Fig.~\ref{fig:2}, with symbols/colors corresponding to 
{  $(\dot\gamma,n)=$ in the 2D:  $:\{5\times10^{-4}/s,5\}$, $\star:\{5\times10^{-4}/s,10\}$, $\bullet:\{5\times10^{-3}/s,5\}$, $:\{5\times10^{-3}/s,10\}$, $\triangleleft:\{5\times10^{-4}/s,10\}$  and also, 3D: $\square:\{5\times10^{-4}/s,10\}$. The 3D sample has identical loading and strain-rate sensitivity as the 2D case $:\{5\times10^{-4}/s,10\}$.}
(f) {  The first component projection, $P^{(k)}_1$,} and in the background the stress-strain curves from Fig.~\ref{fig:2} are shown. A clear correlation with stress-strain behavior, is seen in the peak of the second PCA component and the decrease of the first PCA component, signifying the onset of the elastic-plastic transition.
(g) {  The second component projection, $P^{(k)}_2$,} defined in text, as function of the applied strain, 
 (h) The PCA-predicted yield point Vs, the actually --from the stress-strain curve-- found yield point
 }
\label{fig:4}
\end{figure*}

Regarding the plasticity model, we  consider~\cite{pap2019, asaro, roters2019} the case of  tensile loading in the x-direction, thin-film and bulk specimens which are periodic in all directions, and have sample dimensions in (x,y,z): (512,512,2) or (64,64,64), where each cell can be interpreted as a representative volume element of dimensions 2$\mu$m in each direction,  promoting the perspective of investigating a $1$mm$^2$ thin film or sub-mm bulk samples. {  We study two different geometries so that the geometry effect on our measures is checked.} The crystalline structure of the material is {  face-centered cubic (FCC) Aluminum (Al),} with standard stiffness coefficients $C_{11}=106.75$GPa, $C_{12}=60.41$GPa, $C_{44}=28.34$GPa (in reference to the cubic coordinates). {  The importance of the examples studied is that the simulated dimensions can be achieved by common DIC procedures~\cite{Schreier:2009xd}.}

Details of the model's hardening dynamics can be also found in Refs.~\cite{pap2019, roters2019}. {  The utilized model is built upon commonly used modern phenomenological crystal plasticity assumptions, which captures effectively the essential aspects of slip-based macroscale plasticity through a combination of fundamental assumptions and basic constitutive laws that have been confirmed in various metals over the last four decades~\cite{asaro}. While our modeling examples cannot be accurate in capturing delicate plasticity features that relate to defect-driven latent hardening or non-Schmid effects~\cite{raab}, they still capture in a self-consistent manner the basic physical mechanisms involved in the transition from elasticity to plasticity, as it takes place in most metals. In summary, in the examples studied,} the model captures finite deformations in a cubic grid, which are used to calculate constitutively plastic distortion rates along all 12 FCC slip systems. All samples have 256 grains that are distributed randomly using a Voronoi tesselation~\cite{nr} (see also Fig.~\ref{fig:2}(a)) and the model is solved by using a spectral approach~\cite{pap2019}. 
At this stage, several constitutive plasticity laws are required: For estimating the plastic deformation tensor and its rate, we iteratively solve: 
\bea
\dot{F_p} = L_p F_p
\eea
where $L_p=\sum_\alpha \dot\gamma^\alpha s^\alpha\otimes n^\alpha$.
where s and n being the unit vectors along the slip direction and slip plane normal, respectively, and $\alpha$ is the index of a slip system. The total deformation tensor is commonly splitted in elastic and plastic through $F=F^eF^p$. We consider all 12 FCC slip systems which may become active in our single crystal with fixed crystalline orientation. The slip rate $\dot\gamma^\alpha$ is given by the basic phenomenological crystal plasticity constitutive equation~\cite{asaro,kalid92}, 
\bea
\dot\gamma_\alpha = \dot\gamma_0 \Big|\frac{\tau^\alpha}{g^\alpha}\Big|^n sgn(\tau^\alpha)
\eea
where $\dot\gamma_0=0.001/s$ is the reference shear rate, $\tau^\alpha = S\cdot (s^\alpha \otimes n^\alpha)$ is the resolved shear stress at a slip resistance $g^\alpha$, with $S=[C]E^\epsilon$ being the Second Piola-Kirchoff stress tensor, n is taking the values (5,10) in this work (inverse of the strain rate sensitivity exponent $m=1/n$) and $g^\alpha$ is the slip resistance for a slip system $\alpha$.  Hardening is provided by another saturation-type crystal plasticity constitutive law~\cite{brown}: 
\bea
\dot g^\alpha = \sum_{\beta=1}^{12} h_{\alpha\beta} |\dot\gamma^\beta|
\eea
where $h_{\alpha\beta}$ is the hardening matrix:
\bea
h_{\alpha\beta} = q_{\alpha\beta}h_0\left[sgn\left(1-\frac{g^\beta}{g^\beta_{\infty}}\right)\left|1-\frac{g^\beta}{g^\beta_{\infty}}\right|^{p}\right]
\eea
which phenomenologically captures the micromechanical interactions between different slip systems. Here, $h_0=75MPa$, $p=2.25$, and  are slip hardening parameters, that are typically assumed to be identical for all FCC systems owing to the underlying dislocation reactions. $g^\beta$ is the resistance to shear on the $\beta$ slip system, and $g^{\beta}_\infty$ is the saturated shear resistance on the slip system $\beta$ (set at $g^s=63MPa$ for all slip systems), and the shear resistances asymptotically evolve towards saturation. The parameter $q_{\alpha\beta}$ is a measure for latent hardening and its value is taken as 1.0 for mutually coplanar slip systems, and 1.4 otherwise, rendering the hardening model anisotropic. It is common in the literature to find several variants of the above constitutive laws, nevertheless we believe that the results in this work are relatively model independent.

%\begin{figure}[htb]
%\centering
%\includegraphics[width=0.49\textwidth]{Fig4}
%\caption{{\bf Von Mises Stress VS grain orientations}. Von Mises stress and Bunge Euler Angle orientations cos$\phi_1$cos$\phi_2$-sin$\phi_1$sin$\phi_2$cos$\Phi$, for loading rate and rate sensitivity exponent $(\dot\gamma,n)=$ $\{5\times10^{-4}/s,5\}$ }
%\label{fig:4}
%\end{figure}

A known way to control and modify the yield stress in this model is through the change of the rate sensitivity exponent $m=1/n$. A change in the value of $n$ naturally occurs through changes in conditions (eg. temperature). Here, we modify the exponent $n$ by considering the values 5 and 10, while also the strain rate is modified through 3 orders of magnitude ($5\times10^{-4}$, $5\times10^{-3}$ and $5\times10^{-2}$). As it is clearly seen in Fig.~\ref{fig:2}(d), the yield stress is changing by a factor of 2 and hardening exponent increases drastically as well. Also, each point on Fig.~\ref{fig:2}(d) is implying a corresponding image of total strain that is directly captured during the simulation, emulating the DIC procedure~\cite{pap1,pap2}.
Stress/strain profiles generated in our simulations (Fig.~\ref{fig:3}(a-i)) can be readily resembled to profiles that may be generated by DIC techniques for polycrystalline metals at small, quasi-elastic applied loads~\cite{Schreier:2009xd}.

A first impression, after observing the system behavior, is typical: The Von Mises stress fluctuations across a particular sample (seen in Fig.~\ref{fig:2}(c)) do not display any particular orientation preference across the polycrystalline sample of the case $(\dot\gamma,n)=\{5\times10^{-2}/s,10\}$, at total strain $0.25\%$. Also,  patterning in images such as the ones in Fig.~\ref{fig:3} appears commonly mundane: There appears to be no clear correlation of the impact that yielding has on the total strain, since the plastic distortion $F_p$ becomes relatively important. The principal reasons for this mundane impression are clearly, the presence of an always finite, superposing, plasticity-dependent elastic contribution, and also, the fact that plasticity is not homogeneous in the sample, with only few locations at high values, pointing to the fundamentally important plastic localization physics~\cite{asaro}. These findings are consistent to an accumulated understanding in materials science that grain-boundary misorientations are in larger correspondence to emerging stresses and plastic distortions, than actual grain orientations. Nevertheless, to this day, classification of misorientations in polycrystals represents a daunting classification challenge, where there is a space of five relatively unrestricted misorientation dimensions on a grain boundary~\cite{sethna}. In this paper, we focus on two tools for generating  measures that naturally, and fundamentally, relate to the plasticity onset. At the possible expense of computational complexity, we introduce two total-strain measures that are sensitive to the onset of plasticity: First, one coming from a PCA transform analysis of the total-strain images, designed to capture total strain fluctuations, and the other coming from a DWT analysis, designed to capture the onset of strain localization.

\begin{figure}[tbh]
\centering
\includegraphics[width=0.5\textwidth]{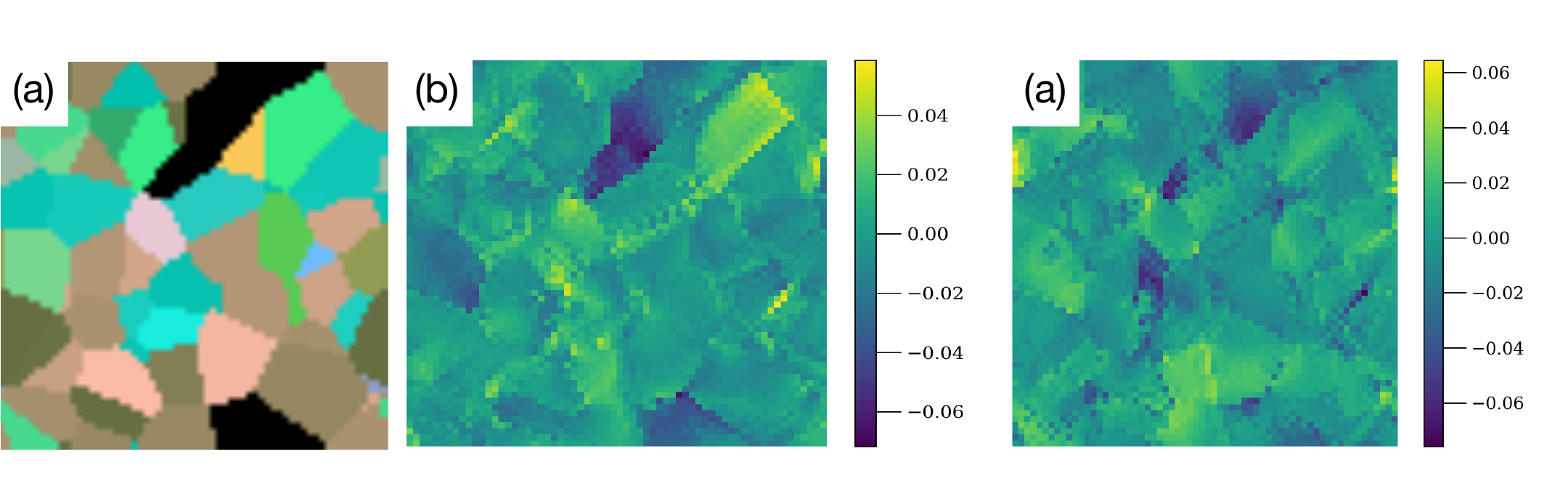}
%\fbox{\rule[-.5cm]{3cm}{3cm} \rule[-.5cm]{3cm}{3cm}}
\caption{{\bf PCA for surfaces of 3D samples}. A 64x64x64 simulation was performed at $(\dot\gamma,n)=$ $:\{5\times10^{-4}/s,10\}$ and the same process as Fig.~\ref{fig:4} leads to analogous results. {  A comparison of the overlap in 2D and 3D shows good agreement, with no qualitative differences.}}
\label{fig:5}
\end{figure}

%Results.
Principal component analysis serves the change of basis in $N$ data vectors $E_k$ (here capturing the norm $|F-I|$ of total distortion across the sample), each entry of which capturing spatial locations of total number $V$, so that predominant correlations are detected in few principal directions. In this case, each $E_k$ is a flattened total-strain vector of $N=L_x*L_y$, corresponding to an $L_x\times L_y$ total-strain image (see Fig.~\ref{fig:1}, Fig.~\ref{fig:3}(c,f,i)). The success of the method is gauged by the capacity to capture, almost completely ($>98\%$), the relative fluctuations in just 2-3 principal vectors, like the one shown in Fig.~\ref{fig:4}(a) or the ones shown in Figs.~\ref{fig:4}(b-c), \ref{fig:5}(b-c)  for our case. In this application, as can be seen in Figs.~\ref{fig:4}(d), PCA works well. Each component (eg. Fig.~\ref{fig:4}(a)) represents a ``characteristic" fluctuation pattern in the evolution of  data vectors. 

 Then, these vectors can be projected back to the original data vectors through a vector dot product, so that overlap with  data vectors is tracked. For the data vectors' fluctuations to be directly comparable, each data vector shall have the same average value ('0') and standard deviation ('1'), so data vectors are rescaled in a standard way:
\bea
\tilde E_k = \frac{E_k - \langle E_k \rangle}{\sqrt{\langle E_{k}^2 \rangle - \langle E_k \rangle^2}}
\eea
where the $\langle\rangle$ imply a spatial average $\langle E_k\rangle = \sum_{i} E^{(i)}_k$ with $i$ counting spatial locations.

Principal components emerge through the singular value decomposition of the matrix $X = \{E_k\}$ for $k\in\{$range where data is collected$\}$, including N lines and $V$ columns and then focusing on the top singular values and components. In other words, the rectangular NxV matrix $X$ {  is decomposed to
\bea
X=U\Sigma W^T
\eea 
where} $U$ is an NxN matrix, containing columns that are orthogonal unit vectors of length N that are known as ``left singular vectors" of $X$. Also, $W$ is a $VxV$ matrix whose columns are also orthogonal unit vectors of length $V$ and are known as the right-singular vectors of $X$. For every singular vector $s_i$, there is a corresponding singular value $\sigma_i$.

The condition of capturing most fluctuations in the system is literally equivalent to identifying the eigenvalues $\sigma$ of the covariance matrix of the data vectors $C=X^T X/(n-1)$, which represent the variance of the fluctuations, and then the top 3 eigenvalues should roughly equal ~$99\%$. It is easy to show that $C=V(\Sigma^2/(n-1))V^{-1}$, which promotes the equivalence of singular values with $\sqrt{\lambda (n-1)}$. Plotting the sum of $\sigma$, as more principal components are added, demonstrates (see Fig.~\ref{fig:4}(d)) that keeping only two components promotes a complete capture of the total data variance across loading it. In retrospect, this is an outcome of the fact that plasticity is literally controlled principally by 2 states: one before yielding, and the other afterwards.

Given that the singular values are labeled $\sigma_i$ for the $i$-th component, the projection (dot product) of a component $s_i$ on one of the data vectors $\tilde{E}_k$ is defined here as: { 
\bea
P^{(k)}_{(i)}=\frac{s_i\cdot E_k}{\sqrt{\sigma_i}}.
\eea 
As seen in Fig.~\ref{fig:4}(e), where $P^{(k)}_1$ and $P^{(k)}_2$ are plotted,
 the plot of the first and second component projections with respect to each other, displays a  consistently similar behavior for every loading case, with the maximum  lying at the sample's yield point. By plotting the projections with respect to the applied strain, it becomes clear that the first component projection is decreased at yielding, while the second has a clearly fluctuation-driven behavior, peaking at the plasticity transition. The identification of the maximum of the 2nd component projection as the yield point can be made through the plot of the $2^{nd}-$component projection with respect to the applied strain. In addition, the yield strain $e_y$ can be defined as the finite extremum of the $2^{nd}-$derivative of stress with respect to the strain, after fitting it with a continuous third-order polynomial. In this way, we can compare the two definitions, shown in Fig.~\ref{fig:4}(h).} So, instead of using an engineering definition of yield stress, we suggest that the yield strength of any sample is uniquely defined as:
\bea
\epsilon_y|_{PCA} &=& {\rm LOC}\{ \frac{dP^{(k)}_2}{d\epsilon_k}=0\} \\
\sigma_y|_{PCA} &=& C \epsilon_y
\eea

The use of the PCA analysis is focused on understanding the \emph{fluctuations} present in the images of total strain. However, another possibility emerges through the investigation of localization of total strain, a characteristic feature of crystal plasticity. This feat can be achieved through using a discrete wavelet transform (DWT) that is designed for capturing localized features in any image. In our case, we consider the image, where x-direction signifies ``space", while y-direction signifies applied-load evolution.

\begin{figure*}[tbh]
\centering
\includegraphics[width=0.99\textwidth]{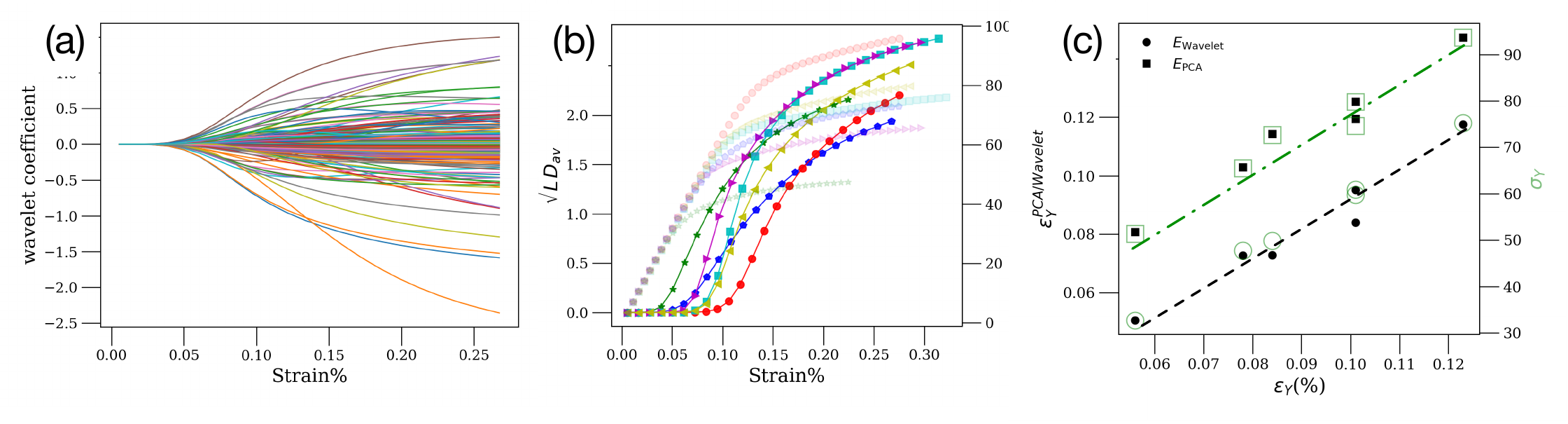}
\caption{{\bf Wavelet Coefficients and the onset of yielding}. {  (a) A subset of de Daubechies coefficients in the strain direction, as function of the applied strain for the case $(\dot\gamma,n)=$ $:\{5\times10^{-4}/s,10\}$, (b) Average, spatially, wavelet coefficients as function of strain for all cases with color/symbol codes analogous to the other figures: {  $(\dot\gamma,n)=$ in the 2D:  $:\{5\times10^{-4}/s,5\}$, $\star:\{5\times10^{-4}/s,10\}$, $\bullet:\{5\times10^{-3}/s,5\}$, $:\{5\times10^{-3}/s,10\}$, $:\{5\times10^{-4}/s,10\}$  and also, 3D: $\square:\{5\times10^{-4}/s,10\}$.}, The average onset of de Daubechies coefficients coincides with the yield point $\epsilon_y$, detected through measuring applied loads. (c) $\epsilon^{Wavelet}_Y$ is defined through the point of the first non-zero value in (b). A natural comparison between PCA and wavelet predictions is shown, for both the yield strain $\epsilon_Y$ and yield stress $\sigma_Y$, where the projected yield stress is just the yield strain multiplied with the elastic modulus. Lines are guide to the eye.}}
\label{fig:6}
\end{figure*}

The discrete wavelet transform (DWT), as formulated by Daubechies~\cite{daubechies1} has been extensively used to decompose timeseries fluctuations. In contrast to Fourier transforms, that decompose fluctuations across a Fourier frequencies' continuum, the wavelet decomposition is performed across a discrete set of scales, and the component amplitude measures how much variability exists between adjacently located averages associated with a particular scale. Examples the wavelet decomposition's usefulness are abundant in science and engineering, and for example, include sleep state patterns~\cite{wl1} and  stochastic fluctuations on binary stars~\cite{wl2} . 

Regarding wavelets, they are functions in the set of real numbers to the set of real numbers, each of which is derived from the mother using translation and scaling: $\Psi_{s,x}(t) = 2^{s/2}\Psi(2^s t + x)$ where: $s, x$ - real numbers, $\Psi$ - mother wavelet, $\Psi_{s,x}$ – wavelet of scale $s$ and translation $x$, in other words
\bea
\Psi^{s,x}(t) = \frac{1}{\sqrt{|s|}}\psi\left(\frac{t-x}{s}\right)
\eea
for $s\neq0$, with $<\Psi>=0$.
In this work, we use the Daubechies wavelet transform, which uses the basic DWT structure, where any square, real, integrable function ($y(t)\in L^2\{R\}$) can be expressed in terms of transform at multiple scales:
\bea
y(t) = \sum_m \sum_n \tilde{y}_{m,n} 2^{-m/2} \bar{\psi}(2^{-m}t - n)
\eea
and
\bea
\tilde{y}_{m,n} = \int_{-\infty}^\infty 2^{-m/2} \psi(2^{-m}t - n)y(t)dt
\eea
while the functions $\psi$ and $\bar{\psi}$ are the wavelet basis functions for analysis and synthesis~\cite{daubechies1}, and they are biorthogonal ($\langle \psi(t)\bar\psi(t-m)\rangle=\delta_{m,0}$ and $\langle \psi(t)\bar\psi(t-m)\rangle=\delta_{m,0}$). 
Towards the generalization of this construction in two dimensions, the 2-D wavelet transform of an image $f(x)$ computes the set of inner products
\bea
d_{j}^k[n] = \langle f, \psi_{j,n}^k \rangle
\eea
for scales $j\in Z$, position $n\in Z^2$ and orientation $k\in\{H,V,D\}$.

The wavelet atoms are defined by scaling and translating three mother atoms $\{\psi^H,\psi^V,\psi^D\}$:
\bea
\psi^{k}_{j,n}(x)=\frac{1}{2^j}\psi^k\left(\frac{x-2^jn}{2^j}\right)
\eea
 These  scaling  function  and  directional  wavelets  are  composed  of the product of a one-dimensional scaling function $\phi$ and corresponding wavelet $\psi$ which are demonstrated as the following: 
 \bea
 \phi(x,y)=\phi(x)\phi(y)\\
 \psi^H(x,y)=\psi(x)\phi(y)\\
 \psi^V(x,y) = \phi(y)\psi(x)\\
 \psi^D(x,y) = \psi(x)\psi(y)
 \eea
A high number of vanishing moments allows to better compress regular parts of the signal. However, increasing the number of vanishing moments also increases the size of the support of the wavelets, which can be problematic in part where the signal is singular (for instance discontinuous). In this work, we focus on the properties of $\psi^V$ {  and thus, $d_{2}^D$.}

Here, our 2D images are composed of total strain features in columns, while each row contains a different applied strain. For clarity, in this way, the images have 30 rows but $2^{18}$ or $2^{16}$ columns. We focus solely on the features of $\psi^V$ which capture localization along the loading direction. By plotting DWT coefficients (just a small sample of 500 coefficients, out of the $2^{18}$ or $2^{16}$ ones) along the loading direction (see Fig.~\ref{fig:6}), for a particular yielding case, one can see that the coefficients become non-zero only at a particular value, which can be associated to the yield point. By performing a spatial average of all coefficients' absolute values, one can {  calculate 
\bea
D_{av}=\frac{1}{N}\sum_{n\in[1,N]} d_{2}^D[n]
\eea
where $N$ is the number of pixels, and the behavior of $D_{av}$, is shown across all cases in Fig.~\ref{fig:6}(b), demonstrating that coefficients acquire non-zero values at a point that resembles very much the yielding transition,} So, in this way, we suggest that another computational definition of the yield point {  is:

\bea
e_y|_{Wavelet} &=& {\rm LOC}\{ max(D_{av}=0)\} \\
\sigma_y|_{Wavelet} &=& C * e_y
\eea
}
%\section{Discussion}
{  The PCA and wavelet yield point definitions can be also compared directly, shown in Fig.~\ref{fig:6}(c).} As the direct comparison shows, the predictions from the two methods are closely related, possibly giving an edge to wavelets if the focus is on the identification of the departure from elasticity, while the predictive edge belongs to PCA when elastoplastic fluctuations become maximum, and elastic deformation disappears. 

It is worth noticing that it was expected that the PCA predictions are always larger in magnitude than the wavelet ones. The reason is that the wavelet-based criterion is concentrated on detecting the onset of localized fluctuations, while the PCA-based one is focused on detecting the maximum of the fluctuations (not localization). Nevertheless, the evidently constant offset between wavelet and PCA predictions seens accidental, due to the fact that elastic-plastic nonlinearity in the models considered is quantitatively similar. A possible consideration of highly non-linear elastic responses and unconventional hardening should have a strong effect on the offset between wavelet and PCA predictions.

In conclusion, we demonstrated that two distinct computational tools, easily implemented through modern computational software, can be used towards the identification of yielding in polycrystals through only total-strain-measurements, without the need to separate elastic from plastic contributions from total strain profiles. 
{  In the simulations performed, each pixel's linear dimension corresponds to $2\mu$m and the total size of tracked surfaces approach 1mm, thus typical experimental tests should be capable of matching the conditions assumed in this work. By comparing two distinct geometries, we found that the transition from elasticity to plasticity is equally detectable on a surface of a 3D crystal, as much as in a 2D thin film. The effects reported here originate in the long-range nature of the interaction between elasticity and plasticity in macroscopic terms, without explicit connection to the character and dynamics of plasticity defects such as dislocations. 
These long-range interactions influence not only the yield point, as shown in this work, but also the hardening and damage sample behaviors beyond the yield point. Characteristically, one may tentatively conclude through data such as in Fig.~\ref{fig:4}(g) that the slope of $P^{(k)}_2$ beyond yielding is inversely related to the hardening coefficient. Through preliminary studies, we find that analogous dependencies may be found in simulations of damaged specimens. The generality of these dependencies, and the formulations towards microscopically motivated  scaling functions~\cite{rev}, will be the subject of future works. In addition, it is worth exploring experimental validation routes and ways to identify further material properties by detailed analysis of total-strain fluctuations in evolving images. 
}

\begin{acknowledgements}
We acknowledge support from the European Union Horizon 2020 research and innovation program under grant agreement no. 857470 and from the European Regional Development Fund via Foundation for Polish Science International Research Agenda PLUS program grant no. MAB PLUS/2018/8. MJA would like to acknowledge the Academy of Finland via the grant 317464. We acknowledge the computational resources provided by the National Centre for Nuclear Research in Poland. The data that support the findings of this study, together with Python analysis scripts can become available from the corresponding author (SP) upon reasonable request.
\end{acknowledgements}

%\bibliography{eqfree} 

\end{document}